\documentclass[reprint]{revtex4-1}

\usepackage[mathlines]{lineno}

\usepackage[T1]{fontenc}
\usepackage[utf8]{inputenc}
\usepackage[english]{babel}
\usepackage{amsmath}
\usepackage{graphicx}
\usepackage{siunitx}
\usepackage{setspace}
\usepackage{upgreek}

\usepackage{color}
\renewcommand{\d}{\textrm{d}}
\newcommand{\LET}{\textrm{LET}}

\newcommand{\Ncol}{N_{\textrm{collected}}}
\newcommand{\Ncre}{N_{\textrm{\red{released}}}}
\newcommand{\Ei}{\textrm{Ei} }
\newcommand{\IT}{\texttt{IonTracks}}
\renewcommand{\pi}{\uppi}
\newcommand{\ea}{\emph{et al.}}

\newcommand{\red}[1]{\textcolor{black}{#1}}

\usepackage{soul} 
\usepackage{todonotes} 

\begin{document}



\title{A general algorithm for calculation of recombination losses in ionization chambers exposed to ion beams} 




\author{Jeppe Brage Christensen}

\affiliation{Department of Physics and Astronomy, Aarhus University, Ny Munkegade 120, Aarhus, Denmark}
\email{jeppebrage@phys.au.dk}

\homepage[Source code: ]{\url{https://github.com/jbrage/IonTracks/releases/tag/1.0}}

\author{Heikki T\"{o}lli}
\affiliation{Department of Radiation Sciences, Ume{\aa} University, SE-901 85, Sweden}

\author{Niels Bassler}
\affiliation{Department of Physics and Astronomy, Aarhus University, Ny Munkegade 120, Aarhus, Denmark}


\date{\today}

\begin{abstract}

Dosimetry with ionization chambers in clinical ion beams for radiation therapy requires correction for recombination effects. However, common radiation protocols discriminate between initial and general recombination and provide no universal correction method for the presence of both recombination types in ion beams of charged particles heavier than protons. Here, we present the open source code IonTracks, where the combined initial and general recombination effects in principle can be predicted for any ion beam with arbitrary particle-energy spectrum and temporal structure. IonTracks uses track structure theory to distribute the charge carriers in ion tracks. The charge carrier movements are governed by a pair of coupled differential equations, based on fundamental physical properties as charge carrier drift, diffusion, and recombination, which are solved numerically while the initial and general charge carrier recombination is computed.

The algorithm is numerically stable and in accordance with experimentally validated theories for initial recombination in heavy ion tracks and general recombination in a proton beam.

IonTracks is validated against the Jaffé's and Boag's theory of recombination in pulsed beams of multiple ion species. IonTracks is able to calculate the correction factor for initial and general recombination losses in parallel-plate ionization chambers. Even if only few experimental data on recombination effects in ionization chambers are available today, the universal concept of IonTracks is not limited to the ions investigated here. Future experimental investigations of recombination in pulsed and possibly also continuous ion beams may be conducted with IonTracks, which ultimately may lead to a more precise prediction of recombination factors in complex radiation fields.
\end{abstract}


\maketitle 

\section{Introduction}

Several chemical and solid-state methods have been developed in the past decades for measuring radiation dose, but ionization chamber dosimetry retains its position as the most common and often most convenient and precise method to determine the absorbed dose in clinical beams. Consequently, dosimetry protocols, such as that from the International Atomic Energy Agency (TRS-398) \cite{IAEA2000}, recommend air-filled ionization chambers for reference dosimetry of therapeutic photon, electron, proton, and ion beams. 

The charge \red{released} in the ionization chamber is related to the absorbed dose. However, the charge collected at the electrodes is less than the charge \red{released} in the ionization chamber, since some ion pairs are lost due to recombination. Several theories for correction of the recombination losses have been proposed in the past: Jaffé \cite{Jaffe1913,Jaffe1929} investigated the \emph{initial recombination}, that is, recombination of charge carriers \red{released} within a single ion track, in gas filled ionization chambers exposed to ion beams. Jaffé deduced a Gaussian shaped charge carrier distribution \red{radial to the primary track} which he applied throughout his work. By taking the mobility of the charge carriers and their geometric distribution in the electrical field into account, Jaffé derived an expression for the ionic collection efficiency, both for the ion tracks parallel and non-parallel to the electric field in the ionization chamber.

Jaffé's work was later extended by Kramers \cite{Kramers1952} for initial recombination in liquid ionization chambers, where charge carrier mobilities are less than those in gas (air). Additionally, since the cavity in the ionization chamber is in condensed phase rather than in gaseous phase, a higher charge carrier density is established. \red{The reduced charge carrier mobilities and higher charge carrier density lead} to a dramatic increase of recombination effects \cite{svenskere}. 

Boag \cite{Boag1950} investigated the recombination in pulsed ion beams, and later developed an extension to the work with Hochhäuser and Balk~\cite{Boag1996} by including a free-electron component. The free-electron component represents the fraction of electrons which escape attachment to molecules, which else would form negative ions. In Thomson's work on conduction of electricity through gases~\cite{Thomson1928}, he formulated the equations that form the basis for \emph{general (volume) recombination} correction, that is, recombination of charge carriers from different ion tracks in continuous beams. Mie~\cite{Mie1904} presented approximate numerical solutions to Thomson's theory on general recombination. These solutions were then applied by Greening~\cite{Greening1964} to formulate the current formalism for general recombination in continuous beams. The initial recombination in electron or photon beams is normally neglected in clinical radiotherapy which leaves only general recombination to be corrected for. The situation is different in ion beams heavier than protons, where both initial and general recombination may occur.  

The TRS-398 protocol provides recommendations for experimental determination of the recombination correction factor for either initial or general recombination in ion beams heavier than protons, and recommends the two voltage method~\cite{Boag1980} to be applied to correct for general recombination in pulsed beams. If the general recombination is negligible, it provides a linear relation to determine the recombination correction factor. Also this linear relation assumes a 1/V dependency. Palmans \ea\ \cite{palmans2006ion} pointed out that the two voltage method for pulsed beams, as described in TRS-398, overestimates the recombination correction with a factor up to 2\% in proton beams. Furthermore, TRS-398 is only explicit for pulsed/scanning beams of ions heavier than protons. 

\red{All heavy ion facilities are synchrotron based whereas only few proton facilities use this technology. The cycle frequency for synchrotrons range from approximately 0.1\,Hz-2\,Hz and the beam itself is bunched with a frequency corresponding to the radio-frequency cavity (order of 100\,MHz). In other words, the fluence is both too slow and too fast to be regarded as a pulsed beam in the context of dosimetry. However, development of faster and stronger accelerators such as Fixed-Field Alternating Gradient (FFAG) \cite{hud15307} and laser induced ion acceleration \cite{karsch2014theoretical} may require special attention to correctly estimate general recombination. }

Even more clinically relevant is intensity modulated particle therapy, where scanning beams may deliver very heterogeneous dose distributions both in space and in particular the time domain. 
General recombination in proton and heavier ion beams may therefore vary depending on the specific treatment plan delivered. High dose-rates with non-negligible general recombination may be seen along the core of the pencil beam when delivering large fields. \red{These large fields require high beam current in order to complete field delivery in short time. Complex multi-field optimized treatment plan validation may consequently require careful assessment of recombination effects as a function of time and position, already stressed by Palmans \ea\ \cite{palmans2010ion} for photon fields used in Tomotherapy.}

\red{2D-ionization chamber arrays are typically employed in clinical settings for complex field validation such as the IBA MatriXX.} \red{However, this particular chamber records at a fixed voltage set by the manufacturer \cite{saminathan2012comparison}, thus it does not provide a possibility to verify the amount of general recombination.} \red{Finally, polarity effects in micro ionization chambers \cite{miller2016polarity}  may require detailed simulations of the recombination effects in complicated electric field configurations.}

\red{Here, the open source software \IT\ for calculation of recombination losses in arbitrary ion beams is presented. \IT\ uses amorphous track structure theory and basic physics to evolve the charge carriers liberated by an incident particle as a function of time, and in principle any beam time structure can be implemented and followed. \IT\ is able to discriminate between initial and general recombination, even in a pulsed mode where both recombination types cannot be separated due to their common 1/V dependence.}

Recombination is a process involving low energy charged particles and cannot currently be corrected satisfactorily for by Monte Carlo calculations. This is due to uncertainties in the cross sections at these low energies, motivating the application of amorphous track structure theory in \IT , which is a more phenomenological approach \cite{AmTrack,AmTrackRadBio}. \IT\ allows an inclusion of a free-electron component to model the effect on recombination of electrons that do not attach to molecules to form negative ions. 

\red{The theoretical aspects are introduced in the following section, where the track structure theory and the movement of charge carriers are outlined. Subsequently, the boundary conditions and methods to solve the partial differential equations, which govern the charge carrier movements, are investigated and discussed. \IT\ is validated against Jaffé's theory for initial recombination, the two voltage method, and finally the inclusion of a free-electron component in \IT\ is compared to a similar extension to the two voltage method.}

\section{Theory}
\subsection{\red{Collection efficiency}}
The Jaffé theory was experimentally validated by Kanai \textit{et al.}~\cite{Kanai1998} for ions heavier than protons. Track parameters based on \cite{Kanai1998} for three ions heavier than hydrogen are given in table~\ref{tab:Kanai_params}, which mainly are used to obtain a substantial amount of initial recombination and to compare \verb"IonTracks" to Jaffé's theory.
The collection efficiency $f$ in \verb"IonTracks" is  calculated as
\begin{equation}
f = \frac{\Ncol}{\Ncre}, \label{eq:collection_efficiency}
\end{equation}
where $\Ncol$ is the number of ions collected at the electrodes and $\Ncre$ is the number of \red{released} (initialised) ions.

\begin{table*}[htbp]
\centering
\caption{\label{tab:Kanai_params} Track parameters used for \IT\ calculations in this paper. LET values are calculated using \texttt{libdEdx} \cite{toftegaard2014improvements}. The corresponding Gaussian track radii $b$ are found in Kanai \ea ~\cite{Kanai1998}.}
\begin{tabular}{l!{\quad}l!{\quad}l!{\quad}l!{\quad}l!{\quad}l}
\hline
 & Unit & Iron ion & Neon ion  & Carbon ion & Proton \\
\hline
Energy & $\si{\mega\electronvolt \per \amu} $ & 40 & 60 & 90 & 100\\
LET (air)	& $\si{\kilo\electronvolt\per\micro\meter}$ & 1.02 & 0.115 	& $3.03\times 10^{-2}$ & $7.76 \times 10^{-4}$ \\
Track radius, $b$		& $\si{\micro\meter}$  & 50 & 20 & 10.5 & 10\\
\hline
\end{tabular}
\end{table*}

\subsection{The charge carrier distribution in the track of a charged particle}
The linear charge carrier density $N_0$, the number of charge carriers per unit length, along the path of the charged particle depends on the linear energy transfer (LET) and \red{the mean energy expended in air per formed ion pair} $W$, and is given as
\begin{equation}
N_0 = \frac{\LET}{W}. \label{eq:linear_density}
\end{equation}
The linear ionization density is assumed to be constant throughout the ionization chamber \cite{boag1966radiation}. Jaffé  modelled the radial charge carrier density $n_0(r)$ in the track of a charged particle as
\begin{equation}
n_0(r) = \frac{N_0}{\pi b^2} \exp \Bigg(- \frac{r^2}{b^2} \Bigg), \label{eq:Gaussian}
\end{equation}
where $b$ is the Gaussian track radius and $r$ is the distance from the centre of the track. \red{Even if all the liberated electrons form negative ions, the distance they move before attachment yields an overall larger track radius for the negative ions, i.e.\ one probably would have to distinguish between positive and negative ion track radii $b_\pm$ with $b_-> b_+$. Consequently, the larger track radius for negative ions leads to a slight decrease in the ionic recombination which however is not the objective of this work. Instead, an average of the two radii is used throughout the work although an actual implementation of different radii in \IT\ is straight forward.}

\IT\ uses the Gaussian charge carrier distribution in the present work, although it should be noted that the algorithm is not limited to any particular type of charge carrier distribution. Kaiser \textit{et al.}~\cite{Kaiser2013} showed that the Gaussian charge carrier distribution matches recombination experiments better than e.g.\ the Scholz-Kraft parametrization~\cite{Scholz1996}. The total number of charge carriers in an ion track is calculated as
\begin{equation}
N_{\text{released}} =  \int_V n_0(r) \, \d V = N_0 \ell, \label{eq:created}
\end{equation}
where $\ell$ is the length of the particle path between the electrodes.

\subsection{Charge carrier movements}
The particle current density $\vec{J}_{\pm}$, in units of charge carriers per unit area per unit time, of both positive and negative ions, consists of contributions from the \red{drift caused by the} electric field $\vec{E}$ and the \red{diffusion caused by the} concentration gradient and is given as
\begin{equation}
\vec{J}_{\pm}(\vec{r},t) = - D_{\pm} \, \nabla n_{\pm} (\vec{r},t) \pm \vec{E} \, \mu_{\pm} \, n_{\pm}(\vec{r},t),
\end{equation}
where $n_{\pm}$ is the charge carrier density, $\mu_{\pm}$ the mobility, and $D_{\pm}$ the diffusion constant for positive and negative ions respectively. One may from the equation of continuity find
\begin{equation}
\frac{\partial n_{\pm} (\vec{r},t)}{\partial t} + \vec{\nabla} \cdot \vec{J}_{\pm} = - \alpha n_+ (\vec{r},t) n_{-}(\vec{r},t), \label{eq:cont_eq}
\end{equation}
where $\alpha$ is the recombination constant and the right-hand side models the recombination of ions \cite{Thomson1899}. Eq.\ (\ref{eq:cont_eq}) is traditionally rewritten as  
\begin{equation}
\frac{\partial n_{\pm} }{\partial t} =  D_{\pm} \, \nabla^2 n_{\pm}\mp \mu_{\pm} \bigg(\vec{E} \cdot \vec{\nabla} n_{\pm} + n_{\pm} \vec{\nabla} \cdot \vec{E} \bigg) - \alpha n_+  n_{-}, \label{eq:theEquation}
\end{equation}
where $ \mu_\pm \vec{E} \cdot \vec{\nabla} n_{\pm}$ is in units of charge carriers per unit volume per unit time 
and ${\mu_{\pm} n_{\pm} \vec{\nabla}\cdot \vec{E}}$ is the change in ion drift velocity caused by space charge screening effects. This effect screens the \red{released} charge carriers from the external electric field, and results in a \red{change} of the observed charge carrier's drift velocity towards the electrode of opposite polarity. \red{The divergence of the electric field is calculated by applying the Maxwell-Gauss equation
\begin{equation*}
\vec{\nabla} \cdot \vec{E} = \frac{q}{\epsilon_0} \big(n_+ - n_-\big) ,
\end{equation*}
where $q$ is the charge of an electron and $\epsilon_0$ is the permittivity.} Consequently, more charge carriers will recombine before they reach the electrodes. The space charge screening term is usually neglected along with initial recombination for low-LET particles such as protons. However, the present inclusion of dense heavy ion tracks increases the space charge screening effect which therefore is included for a more precise study of the recombination process. 

The non-linear partial differential equations in  eq.\ (\ref{eq:theEquation}) govern the movements of the charge carriers. The equations do not have analytical solutions but will be approximately solved by an explicit numeric time integration using the Lax-Wendroff technique \cite{lax1960systems,Press1992,Dehghan2004}. In the case of a free-electron component, the \red{free} electrons are treated as negative charge carriers, but with different mobility and diffusion constant from those of the negative ions.

\begin{table*}[htbp]
\centering
\caption{ \label{tab:params}The positive and negative ion parameters and the recombination constant used for the calculations. The electron drift velocity in air is tabulated for several electric field strengths and pressures in \cite{mcdaniel1964collision}.}
\begin{tabular}{@{}lcll!{\quad}c}
\hline
Constant & Symbol & Unit & Value & Reference\\
\hline
Positive ion diffusion & $D_+$ & $\si{\centi\meter\squared\per\second}$ & $2.82 \times 10^{-2}$ & \cite{Thomson1928} \\
Negative ion diffusion & $D_-$ & $\si{\centi\meter\squared\per\second}$ & $4.35 \times 10^{-2}$ & \cite{Thomson1928}\\
Positive ion mobility & $\mu_+ $ & $\si{\centi\meter \squared \per \volt \per \second}$ & $1.36$ & \cite{Boag1996}\\
Negative ion mobility & $\mu_- $ & $\si{\centi\meter \squared \per \volt \per \second}$ & $2.10 $ & \cite{Boag1996}\\
Recombination constant & $\alpha$ & $\si{\centi\meter\cubed\per\second}$ & $1.60\times 10^{-6}$ & \cite{Kanai1998} \\
\hline
\end{tabular}
\end{table*}

\subsection{The Jaffé theory}
When the ion track is parallel to the electric field, the Jaffé theory for initial recombination \cite{Jaffe1913,Jaffe1929,KanneBearden1936} gives a collection efficiency as
\begin{equation}
f_{\parallel} = \frac{N_{\text{collected}}}{N_{\text{released}}} = \frac{y_1}{y_2}  \exp \left(-y_1 \right) \bigg[\Ei (y_1 + \ln \{1+y_2\}) - \Ei (y_1 ) \bigg]  \label{eq:Jaffe_noangle}
\end{equation}
for 
\begin{equation}
y_1 = \frac{8 \pi D}{\alpha N_0}, \qquad y_2 = \frac{2dD}{\mu b^2 E}, \quad \text{and} \quad  \Ei (x) = - \int_{-x} ^{\infty} \frac{e^{-u}}{u} \d u,
\end{equation}
where $\Ei (x)$ is the exponential integral, \red{$D$ is the averaged positive and negative diffusion constant, similarly with $\mu$ for the positive and negative mobilities}, and $E$ is the electric field strength. The collection efficiency $f_{\theta}$, for an ion track rotated an angle $\theta$ to the electric field, is approximated as \cite{Jaffe1913,Jaffe1929}
\begin{equation}
f_{\theta} = \bigg(1+ y_1^{-1} \mathcal{S}(Z)\bigg)^{-1},  \label{eq:angle}
\end{equation}
with
\begin{equation}
\mathcal{S}(Z) = \exp(Z) \Bigg( \frac{\textrm{i}  \pi }{2} \Bigg) H_0^{(1)} \bigg( \textrm{i} Z \bigg) \quad \textrm{and} \quad  Z =  \Bigg( \frac{\mu b E \sin \theta}{2 D } \Bigg)^2,
\end{equation}
where $H_0^{(1)}$ is the Hankel function of first kind and order zero.

\subsection{The free-electron component}
 Immediately after a charged particle has ionized the medium between the electrodes there will be no negative ions. The negative ions are subsequently formed as the free electrons diffuse and drift and attach to oxygen molecules.
The fraction $p$ of all electrons that remain free from attachment to molecules is \cite{Thomson1928}
\begin{equation}
p = \frac{1}{ad} \left( 1- e^{-ad}\right), \label{eq:free_elec}
\end{equation}
where $a$, in units of inverse length, is a coefficient depending on the gas parameters and the strength of the electric field and $d$ is the electrode gap. For the present simulations, the initial concentration of negative ions $n_-$ and electrons $n_e$ as a function of the distance from the anode ($z=0$) to the cathode ($z=d$) is given by \cite{Boag1996} as 
\begin{eqnarray}
\left. \begin{array}{ll} n_-(z) =  n_+(1- e^{-a z}) \\ n_e(z) = n_+ p \end{array} \right\}  \quad \text{for} \quad 0 \leq z \leq d  , \label{eq:ion_elec_distribution} 
\end{eqnarray}
when an ion track is parallel to the electric field. \red{Thus, the simulation of electron attachment is omitted and an initial homogeneous electron distribution is assumed. This assumption leads to an overestimation of the recombination for $p<1$ as electrons closer to the electrode of opposite sign has the lowest possibility of interaction prior to collection.} \red{However, the electron distribution serves as an upper limit for the recombination of electrons with positive ions and thereby be used to test whether the recombination of electrons with positive ions is negligible as anticipated.}
%
Boag \ea~\cite{Boag1996} proposed three different models to include the free-electron component in eq.\ (\ref{eq:ion_elec_distribution}) \red{into the two voltage method}. 
The results of the three proposed models are:
\begin{eqnarray}
\text{Model 1:} \displaystyle \qquad f_1 &= \displaystyle \frac{1}{u} \ln \Bigg( 1 + \frac{e^{pu} - 1}{p}\Bigg),  \label{eq:model1}\\
\text{Model 2:} \displaystyle \qquad f_2 &= \displaystyle p + \frac{1}{u} \ln \bigg ( 1 + (1-p) u\bigg),  \label{eq:model2}\\
\text{Model 3:} \displaystyle \qquad f_3 &= \displaystyle \lambda + \frac{1}{u} \ln \Bigg(1 + \frac{e^{\lambda (1-\lambda) u }-1}{\lambda} \Bigg)  \label{eq:model3}, 
\end{eqnarray}
for
\begin{equation}
 u = \frac{\alpha}{\mu_+ + \mu_-}\frac{n_+ d^2}{V} \qquad \text{and} \qquad \lambda = 1 - \sqrt{1-p},
\end{equation}
where $V$ is the polarisation voltage of the chamber.

\section{Methods}
\subsection{Implementation} 

\begin{figure*}[htbp]
    \includegraphics[width=1\textwidth]{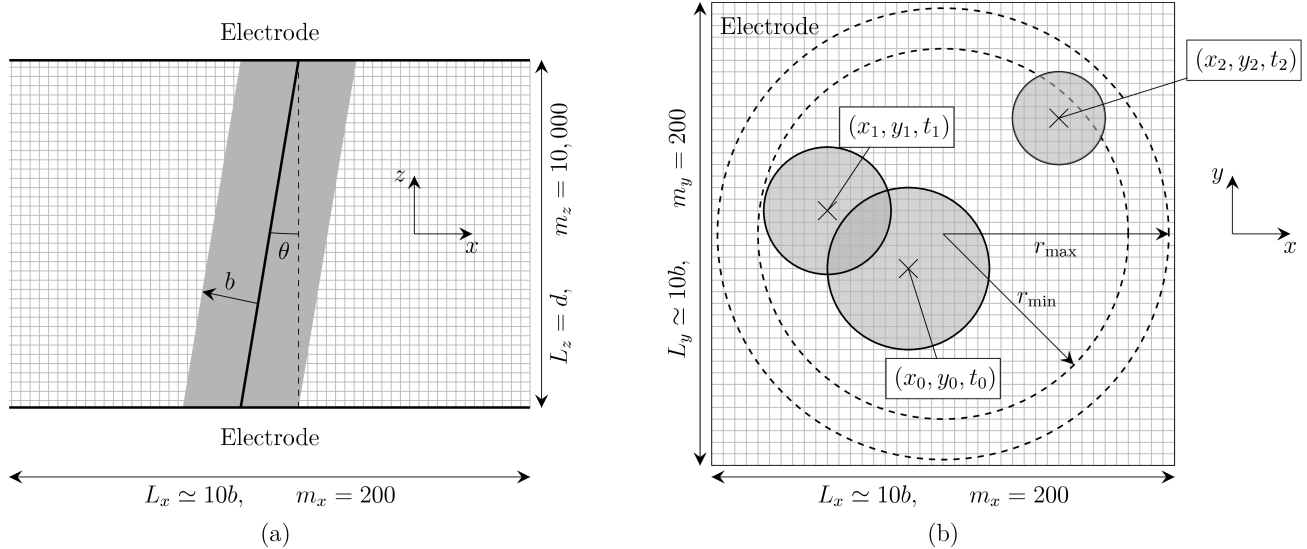}
	\caption{\label{fig:setup} The electrodes are simulated as arrays with widths ${L_x=L_y \simeq 10b}$. $m_i$ denotes the typical number of voxels in the array in the $i$-direction for $i=x,y,z$. (a) The initial setup where a single ion track (track centre marked with a solid line) with Gaussian radius $b$ is rotated \red{by} an angle $\theta \in [0,\pi/2]$ to the electric field between the electrodes with spacing $d$. (b) The setup for the simulation of a continuous beam where three ion tracks have been initialised at different spatial positions ($x,y$) and times $t$.  }	
\end{figure*}

\red{The positive and negative ions are distributed as in eq.~(\ref{eq:Gaussian}) unless a free-electron component is present in which case the negative ions and electrons are distributed as in eq.~(\ref{eq:ion_elec_distribution}).} The recombination of charge carriers of opposite sign, resulting in a loss of charge, is counted continuously throughout the simulation. \red{The present work deals mainly with pulsed radiation, where the pulse duration is short compared to the time it takes to collect the charge carriers, as given in \cite{Boag1996}, and the ionization of the medium between the electrodes by a charged particle is modelled as instantaneous.} 
\red{The collection of charge carriers at the electrode of same sign due to diffusion is included in \IT . However, this effect is negligible since the charge carrier gradient in  eq.\ (\ref{eq:theEquation}) only is directly opposite to the electric field if the particle trajectories are parallel to the electrodes.}
The initial setup is sketched in figure~\ref{fig:setup}(a) where an ion has ionized the medium between the electrodes at an angle $\theta \in [0,\pi/2]$, relative to the electric field, in the solution domain \red{defined by the voxelized volume} ${0\leq x \leq L_x}$, ${0 \leq y \leq L_y}$, and ${0\leq z \leq L_z}$. The spatial dimensions $L_x$ and $L_y$ are chosen large enough to ensure that no significant amount of charge carriers initialised in the array can drift out of the array in a direction perpendicular to the electric field. $L_z$ corresponds to the length of the electrode gap. The charge carriers are removed from the array as they reach either of the electrodes ($z=0$ and $z=L_z$) corresponding to charge collection. Thus, the boundary conditions are
\begin{eqnarray}
n_\pm (x,y,z) = \left\{\begin{array}{ccl} n_\pm(x,y,z) & \text{if} &\quad 0 \leq x \leq L_x,  0 \leq y \leq L_y, \\ &  &   \quad 0 < z < L_z \\ 0 & \text{if} & \quad z=0 \quad \text{or} \quad z=L_z \end{array} \right.
%
\end{eqnarray}
for all times. The initial charge carrier distributions are evolved in time and space on the defined solution domain, as governed by eq.~(\ref{eq:theEquation}), by applying the spatial and temporal derivatives in the appendix. 

The explicit Lax-Wendroff scheme, where a subsequent time step is directly evaluated at the current time step, is in this case computationally faster than an implicit scheme as the Crank-Nicolson \cite{crank1947practical}, where a subsequent time step is found by solving an equation. The superiority in speed is advantageous when dealing with a huge number of computations as in the present case.

For a direct comparison with the Jaffé theory of recombination \cite{Jaffe1913,Jaffe1929}, a single ion track is initialised in the array as given by eq.~(\ref{eq:Gaussian}). The ion track is either initialised parallel or rotated an angle $\theta \in [0,\pi/2]$ to the electric field. The algorithm is implemented in \verb"Python" while computationally demanding routines are converted to the \verb"C" based \verb"Cython" using \cite{Dakin:2015uka} which reduces the computation time with orders of magnitude. The simulation of a single ion track with a space gap of 2\,mm takes less than a minute \red{single} threaded on a laptop with \red{an Intel i5 processor, 1.7\,GHz, and 8\,GB RAM} whereas \red{simulations of} pulsed or continuous beams require several minutes.

\subsubsection{Implementation of the free-electron models.} 
The electron drift velocity is tabulated in \cite{mcdaniel1964collision} for different electric field strengths. Due to a lack of experimentally determined values of the diffusion constant $D_e$ for electrons, the values are estimated from the Einstein relation
\begin{equation}
D_e = \frac{\mu_e k_{\text{B}} T}{q},
\end{equation}
where $\mu_e$ is the mobility of electrons calculated from the drift velocity and corresponding electric field, $k_{\text{B}}$ is Boltzmann's constant, $T$ the absolute temperature, and $q$ the electric charge. The algorithm is modified slightly in order to include electron movements:
\begin{enumerate}
\item The time step $\Delta t$ is first computed to be in accordance with von Neumann's criterion \red{(see appendix)} for the electrons and the simulation is started. During the time it takes for the electrons to be completely collected, the ions will not move significantly and basically no charge carriers will recombine.
\item When the electrons have been collected, the time step $\Delta t$ is re-calculated for the positive and negative ions and the simulation is continued.  The new time step is orders of magnitudes larger than that used for the electron movements.
\end{enumerate}
Generally, the simulation is stopped as the charge carrier distributions of opposite sign have been dragged completely apart.

\subsubsection{Distribution of ion tracks} 
In \IT, a pulsed beam may be considered to be either instantaneous or to consist of short continuous beam segments of any shape.
The ion tracks in such a continuous beam segment are stochastic on the microscopic time scale, and Poisson distributed with an expectation value given by the fluence-rate $\dot{\Phi}$. The fluence-rate is modelled as the product of a spatial component, the fluence $\Phi$, and a time-dependent component $g(t)$ (in units of $\si{\per\second}$), i.e.\ ${\dot{\Phi} = \Phi \, g(t)}$. The surface area of a simulated ionization chamber is typically of the order of ${L_x \times L_y \simeq (0.25\,\si{\milli\meter})^2}$ since a simulation of the entire chamber is too computationally demanding with our current resources. The ionization chamber is assumed to be uniformly irradiated which allows the results obtained in the chosen section of the ionization chamber to be transferred to the entire chamber. 
\red{The function $g(t)$ serves as the mean value for a Poission distribution where the time interval between two consecutive initializations of ion tracks is sampled from.}
\red{Thus, $\dot{\Phi}(t)$ and $g(t)$ are both non-stochastic expectation values used for sampling stochastic ion tracks deposited in the simulation.} 
The ion tracks are distributed uniformly inside a circle with radius $r_{\text{min}}$ in the $xy$-plane as sketched in figure~\ref{fig:setup}(b). \red{The positions of the track centres are sampled uniformly inside the circle with radius $r_{\text{min}}$ corresponding to a uniform irradiation.} A larger radius $r_{\text{max}}$ defines the electrode widths $L_x=L_y = 2 r_{\text{max}}$. The outer radius ensures that no ions are lost due to diffusion in a direction perpendicular to the electric field before they have been collected at the electrodes.

An example of the simulation \red{of a time segment} of a continuous beam is sketched in figure~\ref{fig:setup}(b) where three ion tracks have been initialised at different times and at random spatial coordinates. The charge carriers in the track initialised at time $t=0$ have had time to diffuse further than those in the track initialised at time $t_1$, and general recombination occurs in the overlapping region between the two tracks. Conversely, only initial recombination occurs in the ion track initialised at time $t_2$.

\subsection{Stability and accuracy}
\begin{figure*}[htbp]
    \includegraphics[width=0.75\textwidth]{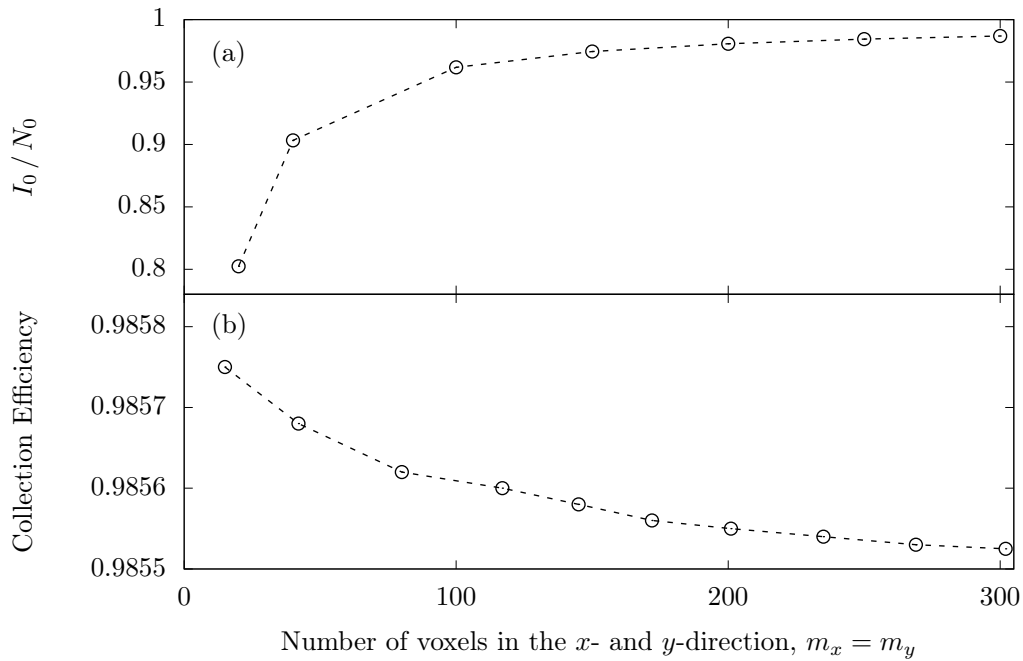}
	\caption{\label{fig:stability} Investigation of the numerical accuracy for fixed array widths $L_x=L_y \simeq 10 b$. (a) shows the ratio of the number of initialised charge carriers to the actual number of charge carriers as a function of an increasing grid resolution. Equivalently, (b) shows the collection efficiency as a function of the grid resolution. The results were obtained with an electric field of $2000\, \si{\volt\per\centi\meter}$, the neon ion track in table~\ref{tab:Kanai_params}, and an electrode gap $d=2\,\si{\milli\meter}$.}
\end{figure*}

The numerical precision of the algorithm is investigated by improving the grid resolution until a satisfactory convergence is obtained, based on the two requirements below:
\begin{enumerate}
\item[(i)] If $N_0$ in eq.~(\ref{eq:linear_density}) denotes the linear charge carrier density along the $z$-direction, i.e. $\theta=0$ in figure~\ref{fig:setup}(a), and $n(x,y,z)$ as in eq.~(\ref{eq:Gaussian}) the charge carrier density, then
\begin{equation}
I_0  \equiv \int_0^{L_x} \int_0^{L_y} n_0(x,y,z) \,\textrm{d} y \, \textrm{d} x \label{eq:unity}
\end{equation}
must equal the linear charge carrier density $N_0$ at $t=0$ for conservation of charge carriers. The integrals are evaluated with Simpson's rule as implemented in the \verb"Python" module of same name.

\item[(ii)] The collection efficiency in eq.~(\ref{eq:collection_efficiency}) of a given initial configuration must converge to the same result if the simulation is repeated with the exact same initial and boundary conditions and ion track distributions. \red{The collection efficiency as a function of decreasing distance between two neighbouring voxels is investigated in order to find a suitable grid spacing.}
\end{enumerate}
The two cases are studied in figure~\ref{fig:stability} for fixed array widths $L_x=L_y\simeq 10b$ using the neon ion track parameters in table~\ref{tab:Kanai_params}. Some charge carriers are lost in the array as the Gaussian distribution is truncated at some point, hence $I_0/N_0 < 1$ as seen in figure~\ref{fig:stability}(a).  At $m_x=m_y=200$, roughly $2\%$ of the charge carriers are not initialised in the array but an inspection of the charge carrier distribution revealed that the majority of the lost charge carriers were missing at the Gaussian tails. Since recombination is a process that depends on the product of the positive and negative charge carrier densities, the effect of the lacking charge carriers at the Gaussian tails thus is negligible. The collection efficiency calculated with \verb"IonTracks" converges to the same value if the simulation is repeated with the same initial and boundary conditions, and henceforth a grid spacing corresponding to $\Delta x = \Delta y = \Delta z = L_x/m_x = b/20$ is applied.

\section{Results and discussion}

\subsection{Comparison with the Jaffé theory}
\begin{figure*}[htbp]
    \includegraphics[width=1\textwidth]{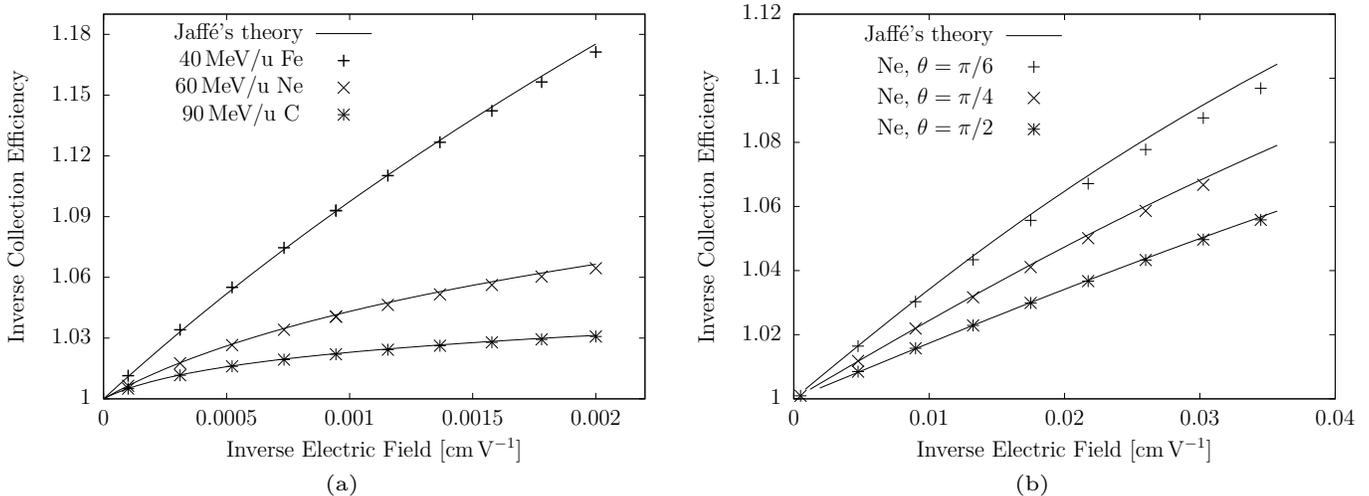}
    \caption{\label{fig:jaffe} Jaffé's theoretical collection efficiencies eq.~(\ref{eq:Jaffe_noangle}) and eq.~(\ref{eq:angle}) compared to the results obtained with \texttt{IonTracks} for three different ion tracks, i.e.\ solely initial recombination in a single track, with track parameters given in table~\ref{tab:Kanai_params}. (a) The iron, neon, and carbon ion tracks parallel to the electric field and (b) the neon ion track rotated an angle $\theta$ to the electric field.}
\end{figure*}

\verb"IonTracks" is used to calculate the collection efficiency for the three heavy ion tracks in table~\ref{tab:Kanai_params} at different electric field strengths. The simulation parameters are listed in table~\ref{tab:params} and the electrode spacing was $2\, \si{\milli\meter}$. The case of ion tracks parallel to the electric field is shown in figure~\ref{fig:jaffe}(a) where an excellent agreement with Jaffé's theory eq.~(\ref{eq:Jaffe_noangle}) is seen for all three types of heavy ion tracks, i.e.\ different LET's and corresponding track radii. The situation of the 60\,MeV/u neon ion track rotated by an angle ${\theta \in [\pi/6,\pi/2]}$ to the electric field is shown in figure~\ref{fig:jaffe}(b). Here, \verb"IonTracks" is in accordance with Jaffé's theory as well, but the discrepancy is increasing with decreasing angle, which is attributed to the fact that the approximated collection efficiency in eq.~(\ref{eq:angle}) only is valid for large angles. The overall agreement between Jaffé's theory and \verb"IonTracks" implies that the algorithm correctly calculates the collection efficiency for initial recombination.

\subsection{Comparison with Boag's free-electron models in pulsed beams} 
\begin{figure*}[htbp]
\includegraphics[width=1\textwidth]{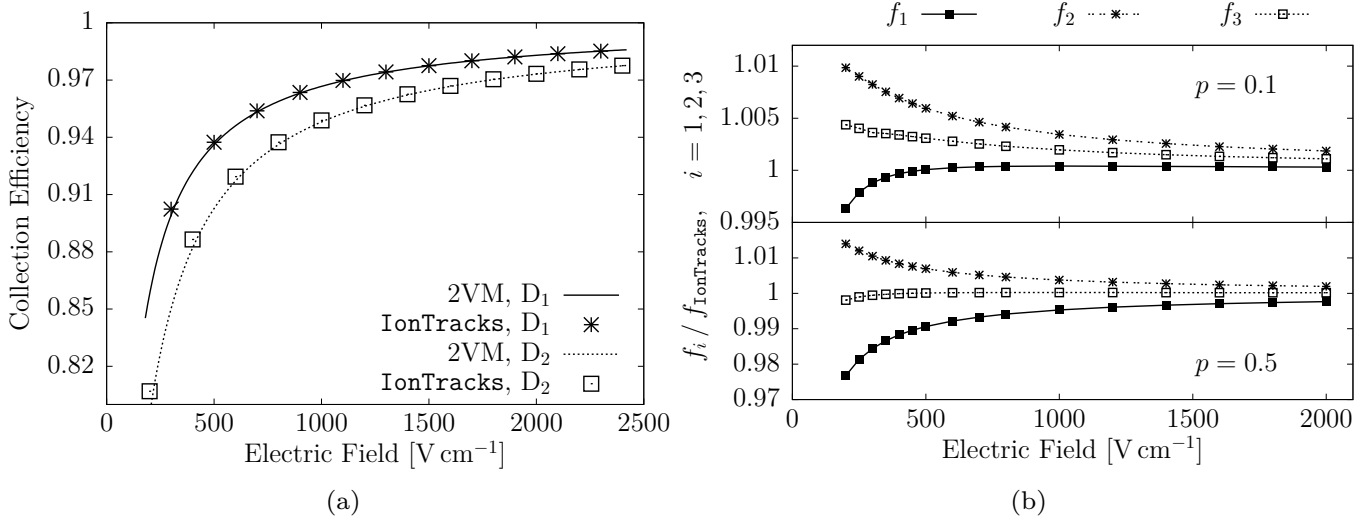}
\caption{\label{fig:free_elec}(a) Comparison of \texttt{IonTracks} with the two voltage method (2VM) for a pulsed proton beam at two doses $D_2 > D_1$. (b) The ratio of the collection efficiency $f_i$ from Boag's free-electron model $i$ for $i=1,2,3$ to the collection efficiency obtained with \texttt{IonTracks}, $f_{\texttt{IonTracks}}$, for free-electron ratios of $p=0.1$ and $p=0.5$. The dose $D_1$ from (a) was applied. }
\end{figure*}

The collection efficiencies in a 100\,MeV proton beam at different electric field strengths were calculated with \verb"IonTracks" for two different doses $D_2>D_1$, where the entire dose was delivered in an instantaneous pulse. Calculations with \IT\ confirm the negligible amount of initial recombination ($<0.1\%$) due to the low LET of a 100\,MeV proton. The results are compared to those from Boag's free-electron models 1, 2, and 3 in figure~\ref{fig:free_elec}(a) for $p=0$, for which all three models \red{converge to the two voltage method $f = u^{-1} \ln (1+u)$}. The convincing accordance between theory and \verb"IonTracks" implies that the simulation of general recombination works correctly. Boag's models for negative charge carrier distributions are in figure~\ref{fig:free_elec}(b) compared to \verb"IonTracks" for $p=0.1$ and $p=0.5$. For $p=0.1$ model 1 seems to be the most appropriate model and all models converge toward the results obtained with \verb"IonTracks" as the electric field is increased, i.e.\ the recombination is decreased. \red{The recombination between the free electrons and positive ions in \IT\ was found to be completely vanishing ($<10$ ppm) and the recombination of charge carriers is only due to the recombination between positive and negative ions.} 

The negative charge carrier distributions in the three Boag models all differ from the distribution they have in \IT, i.e.\ eq.~(\ref{eq:ion_elec_distribution}), and the results in figure~\ref{fig:free_elec}(b) consequently deviate from unity. For low electric fields this discrepancy is increased. For larger fractions of $p$, model 3 fits the actual distribution of negative charge carriers best, which also is reflected in figure~\ref{fig:free_elec}(b) for $p=0.5$. This agrees with the conclusions obtained by Boag \ea\ in \cite{Boag1996}. 

\section{Conclusions}
\verb"IonTracks" is numerically stable and in excellent agreement with the Jaffé theory for initial recombination. Jaffé's theory has been experimentally validated \cite{Kanai1998,KanneBearden1936}, which leads to the conclusion that \verb"IonTracks" correctly simulates the initial recombination in heavy ion tracks. Similarly, the simulation of a 100\,MeV pulsed proton beam permits a comparison between \verb"IonTracks" and the two voltage method at two doses. The good agreement at both doses leads to the conclusion that also general recombination is simulated correctly. Furthermore, \red{three models to include a free-electron component into the }two voltage method were investigated, where the best agreement was found between \verb"IonTracks" and model 3, which is in accordance with the conclusion in \cite{Boag1996}. Space charge screening of the collection field is included in \verb"IonTracks". The agreement with the two voltage method, which does not take space charge into account, implies that the space charge screening term can be neglected in low-LET radiation such as protons, also concluded by \cite{Thomson1928,Boag1996} for photons. 

\verb"IonTracks" currently is extended to include particle energy spectra from FLUKA~\cite{fluka} and SHIELD-HIT12A~\cite{SHIELDHIT} in order to simulate the recombination in particle-energy spectra as e.g.\ found in degraded or mixed ion beams. Furthermore, the algorithm can quantify for what heavy charged particle fluences (or fluence-rates) the recombination makes a transition from being solely initial to a non-linear combination of initial and general recombination, that is, the transition where track overlap is non-negligible anymore.
Given the dose-rate and particle-energy spectrum, \IT\ has the ability to quickly calculate recombination correction factor in ion beams.

\begin{acknowledgments}
The authors wish to thank Prof.\ Bjarne Thomsen at the Department of Physics and Astronomy, Aarhus University, for providing access to his computer cluster.
\end{acknowledgments}

\subsection*{Conflict of interest}
The authors have no relevant conflicts of interest to disclose.

\appendix
\section*{Appendix: Numerical scheme}\label{sec:LaxWendroff}
Let the solution domain be an array with spatial dimensions $L_x$, $L_y$, and $L_z$ and the time it takes to collect all charge carriers $L_t$. The solution domain consists of a mesh of grid-lines
\begin{eqnarray}
x_i &= i \Delta x, \qquad i =0,1,\ldots,m_x=L_x/\Delta x, \\
y_j &= j \Delta y, \qquad j =0,1,\ldots,m_y=L_y/\Delta y, \\
z_k &= k \Delta z, \qquad k =0,1,\ldots,m_z=L_z/\Delta z, \\
t_l &= l \Delta t, \qquad l =0,1,\ldots,m_t=L_t/\Delta t, 
\end{eqnarray}
where $m_x$, $m_y$, $m_z$ are the number of voxels in the $x$-, $y$-, $z$-directions respectively, and $m_t$ the number of time steps, given constant grid spacings $\Delta x, \Delta y$, $\Delta z$ and time step $\Delta t$. The charge carrier density $n(x_i,y_j,z_k,t_l)$ is approximated as $n_{i,j,k}^l$ at the point of intersection of the grid-lines $x_i,y_j,z_k,t_l$, i.e.\ the grid-point $(i,j,k,l)$. Let
\begin{equation}
c_x = \mu_\pm E_x \frac{\Delta t}{\Delta x} \quad  \text{and} \quad  s_x= D_\pm \frac{\Delta t}{(\Delta x)^2},
\end{equation}
be weights, where $E_x$ is the magnitude of the electric field in the $x$-direction, with similar weights in the $y$- and $z$-directions. The derivatives in the Lax-Wendroff scheme \cite{lax1960systems,Press1992,Dehghan2004} are then given as
\begin{equation}
\frac{\partial n}{\partial t} \bigg\vert ^l_{i,j,k} \simeq \frac{n_{i,j,k}^{l+1}-n_{i,j,k}^{l}}{\Delta t}, 
\end{equation}
\begin{equation}
\frac{\partial n}{\partial x} \bigg\vert ^l_{i,j,k} \simeq c_x \frac{(n_{i,j,k}^{l}-n_{i-1,j,k}^{l})}{\Delta x} + (1-c_x)  \frac{(n_{i+1,j,k}^{l}-n_{i-1,j,k}^{l})}{2 \Delta x},
\end{equation}
\begin{equation}
\frac{\partial ^2 n}{\partial x^2} \bigg\vert ^l_{i,j,k} \simeq \frac{(n_{i+1,j,k}^{l}-2n_{i,j,k}^l+n_{i-1,j,k}^{l})}{(\Delta x)^2},
\end{equation}
and similarly in the $y$- and $z$-directions. The largest applicable time step $\Delta t$ is found by computing the largest time step fulfilling the (for positive and negative charge carriers) four equations
\begin{equation}
2(s_x+s_y+s_z) + c_x^2 + c_y^2 + c_z ^2 \leq 1 \quad \text{and} \quad c_x^2c_y^2c_z^2 \leq 8 s_x s_y s_z. \label{eq:vNeum}
\end{equation}
However, a time step $\Delta t$ fulfilling the equations in eq.~(\ref{eq:vNeum}) is only applicable if simulations with identical initial and boundary conditions and track distributions always converge to the same result.


\end{document}